# Electrochemical properties of solid oxide fuel cells under the coupling effect of airflow pattern and airflow velocity*

WANG Hao[1], XIE Jiamiao[1], HAO Wenqian[1], LI Jingyang[1,2], ZHANG Peng[3], MA Xiaofan[1], LIU Fu[1], WANG Xu[1]

1.School of Aeronautics and Astronautics, North University of China, Taiyuan 030051, China

2.Beijing Tsing Aero Armament Technology Co., Ltd., Beijing 102100, China

3.School of Mechanical and Electrical Engineering, North University of China, Taiyuan 030051, China

**Abstract:** Under the dual background of deep adjustment of global energy pattern and severe challenges of environmental problems, solid oxide fuel cell (SOFC) has become the focus of research on efficient and clean energy conversion technology due to its many excellent characteristics. The electrochemical performance of SOFC is affected by various factors such as gas flow pattern (co-flow, counter-flow, cross-flow), flow rate (cathode and anode channel gases), and operating voltage. Accurately analysing the variation of electrochemical indexes with each factor is the basis for proposing the design scheme of high efficiency reaction of the cell. Therefore, a three-dimensional multi-field coupling model of SOFC is established in this study, and the model parameters and boundary conditions covering electrochemistry, gas flow, substance diffusion, etc. are set to study the influence of the coupling between factors on the electrochemical performance of the cell. These results show that with the decrease of operating voltage, the electrochemical reaction rate of the cell increases significantly, the gas mole fraction gradient increases, and the inhomogeneity of the electrolyte current density distribution is enhanced. Under low-voltage operating conditions, the cross-flow flow pattern shows better electrochemical performance advantages, and its power density profile takes the lead in different current density intervals. With the increase of the flow rate of the flow channel gas, the output power density curve of the cell shows an overall upward trend, and then the driving effect of the flow rate increase on the power density increase is gradually weakened due to the saturated cathodic reaction. This study reveals the influence of the coupling of flow pattern, flow rate and voltage on the electrochemical performance of SOFC, and provides guidance for the commercial application of SOFC.

**Keywords:** solid oxide fuel cells, airflow pattern, multi-field coupling, electrochemical performance

## 1. Introduction

Solid oxide fuel cell (SOFC), as the fourth generation of fuel cell technology, directly converts chemical energy into electrical energy in an efficient and clean energy conversion way. It has the remarkable advantages of high specific energy, high efficiency, low noise, low pollution, flexible fuel selection and so on. It shows excellent performance and broad application prospects. In terms of application fields, SOFC can not only be widely used in fixed equipment fields such as fixed power stations, distributed household power supply and emergency power supply, but also be extended to mobile facilities such as surface ships, underwater weapons, unmanned platforms, aerospace, individual power supply and mobile power supply[1–6]. Therefore, SOFC is one of the most promising new energy sources.

The minimum repeating unit of SOFC includes electrolyte layer, anode layer, cathode layer and related connecting components, in which the flow channel is a gas flow channel made of bipolar plate by a specific process. The flow channel is responsible for delivering the gas to the reaction electrode stably and uniformly, and discharging the waste gas generated by the reaction in time. Its structure and characteristics directly affect the performance and life of SOFC. SOFCs generally operate under three different gas flow conditions: co-current, counter-current, and cross-current. In the condition of forward flow and reverse flow, the flow directions of the two gases are parallel, and the flow directions are the same (forward flow) or opposite (reverse flow). In the cross flow condition, the gas flow field is orthogonally distributed. The configuration of flow channels and the gas flow rate have a great influence on the electrochemical performance of SOFC. Different flow patterns will significantly change the gas flow pattern, heat and mass transfer process, and the uniformity of chemical reaction in SOFC. The downstream flow can optimize the temperature distribution under specific conditions and improve the stability of the battery; Countercurrent may promote the full mixing of fuel and oxidant and improve the reaction efficiency; Cross-flow can combine the advantages of the former two, providing a new way to improve battery performance. Through in-depth study of the effects of different flow patterns on the electrochemical performance of SOFC, the flow channel structure of the cell can be accurately designed and the operating conditions can be optimized, so as to improve the overall performance of SOFC and lay the foundation for its wide application in different scenarios. At the same time, it is also very important to study the electrochemical performance of SOFC at different gas flow rates. If the flow rate is too high, the residence time of the gas in the battery will be too short and the reaction will be incomplete; If the flow rate is too low, the supply of reactants will be insufficient, the products will accumulate, the reaction will be hindered, and the performance of the battery will be reduced. Proper gas flow rate can ensure that the fuel and oxidant are uniformly distributed on the electrode surface, promote the timely transport of reactants to the reaction site, improve the electrochemical reaction rate, and enhance the output power of the cell.

  The effects of flow pattern and flow rate on the electrochemical performance of SOFC, including the local current density, temperature and material distribution in the cell, have been explored in many ways[7–12]. Based on the zigzag flow field structure of multi-channel fuel cell, Lu et[13] analyzed the multi-physical field coupling (including velocity, temperature, current density field) and electrochemical performance by improving the external and internal flow field design, and found that the maximum hydrogen flow rate was increased by 7. 83 times compared with the double-outlet design. However, the cross channel enhances the electrochemical reaction and causes polarization loss, which hinders the internal heat dissipation of the cell, resulting in a maximum temperature increase of 71.77 K for the cross channel compared with the double-outlet channel. Based on the kinetic models of SOFC with co-flow, counter-flow and cross-flow configurations, Zhang et al. Obtained the local distributions of temperature, mass flow rate, current density, reactant and product gas concentrations on both sides of fuel and air[14]. It is found that the effect of forward flow and reverse flow is better than that of cross flow, and there is a significant temperature gradient along the length of the cell for the three flow modes. Based on the three-dimensional SOFC model with concurrent gas flow, Andersson et al.[15] that along the direction of gas flow, there is a strong coupling between the local current density and the local temperature, and the heat released by the electrochemical reaction increases the temperature inside the SOFC and accelerates the increase of the current density. Sohn et al. Studied the SOFC with 30% pre-reformed gas as fuel[16], and the results showed that the SOFC with counterflow mode showed better electrochemical performance between the two modes of cocurrent and counterflow. Wang et al.[17] used ANSYS/CFX software to establish a three-dimensional model of planar SOFC, and obtained the distribution of gas mole fraction, temperature, current density and overpotential in the cell. In addition, based on Darcy's law, the flow pattern of gas in the porous electrode region was explored, and the results show that the SOFC with co-current configuration has a higher average current density and a more uniform temperature distribution. The SOFC studied by Choudhary et al.[18] uses syngas (21% methane, 18% carbon dioxide, 20% carbon monoxide, 40% hydrogen and 1% nitrogen) as fuel. The research results show that the average current density of the countercurrent SOFC is higher than that of the concurrent SOFC; However, in terms of efficiency, response time, thermal stress and carbon deposition, the downstream mode has significant advantages, with higher efficiency, shorter response time and excellent performance in dealing with thermal stress and carbon deposition. However, for the SOFC with hydrogen as fuel, the electrochemical performance of the countercurrent SOFC is better than that of the forward flow SOFC, the countercurrent SOFC and the crossflow SOFC[19]. Although increasing the flow rate of air and fuel theoretically helps to improve the performance of the cell, the research results of Sembler et al.[20] show that the cooling effect caused by the increase of flow rate weakens the effect of improving the performance of the cell caused by the increase of flow rate.

To sum up, the current research on the electrochemical performance of SOFC mainly focuses on a single physical condition, such as a single gas flow pattern, a single voltage or a single flow rate, and the research on the influence of different flow patterns on the electrochemical performance of SOFC under different voltages and different gas flow rates is still not comprehensive. Based on this, COMSOL Multiphysics finite element simulation software was used to study the electrochemical performance of SOFC under three flow patterns: forward flow, reverse flow and cross flow. To reveal the influence of gas flow pattern, working voltage and gas flow rate on the electrochemical performance of SOFC, to study the gas mole fraction distribution and electrolyte current density variation in the gas channel under different working voltages, to analyze the polarization curve, power density curve and gas component distribution of SOFC under different gas flow patterns, and to reveal the influence mechanism of SOFC electrochemical performance under different gas flow rates, so as to provide theoretical guidance and data support for further optimizing the electrochemical performance design of SOFC.

## 2. SOFC multi-field coupling model

In the working process of SOFC, electrochemical reaction, mass transfer, momentum transfer and charge transfer are coupled. The electrochemical reaction rate depends on the temperature, the concentration of the substance, and the active surface area of the catalytic reaction; A chemical reaction produces and consumes heat, i.e., the temperature distribution depends on the rate of the chemical reaction and the properties of the solid and gas; The nature of the gas flow depends on the temperature and the concentration of the substance. Therefore, in the theoretical modeling of SOFC, it is necessary to couple the electrochemical model, the gas flow model and the species diffusion model and then solve them jointly.

2.1 Electrochemical model

The rate of an electrochemical reaction can be described as the reaction rate relative to the activation overpotential. Butler-Volmer electrochemical reaction kinetic equation describes the relationship between current and voltage difference at the electrode-electrolyte interface. Based on the Butler-Volmer equation, the SOFC local current distribution[21] can be obtained as follows:

$$i_v = i_0 A_v \left[ \exp\left(\frac{\alpha_a F \eta_{act}}{RT}\right) - \exp\left(\frac{-\alpha_c F \eta_{act}}{RT}\right) \right] , \qquad (1)$$

Where $i_v$ is the specific volumetric current density (A/m²), $i_0$ is the exchange current density, $A_v$ is the active specific area of the porous catalyst layer, $F = 96485$ C/mol is the Faraday constant, $R$ is the gas constant, $T$ is the absolute temperature, $\alpha_a$ and $\alpha_c$ are the SOFC anode and cathode transfer coefficients, respectively, and $\eta_{act}$ is the activation overpotential, which is the difference between the electrode potential and the equilibrium potential.

The concentration overpotential $\eta_{con}$ at the anode and cathode of the SOFC can be calculated as[22] from the following equation:

$$\eta_{con,a} = \frac{RT}{2F} \ln\left( \frac{p_{H_2,b}}{p_{H_2,TPB}} \frac{p_{H_2O,TPB}}{p_{H_2O,b}} \right) , \qquad (2)$$

$$\eta_{con,c} = \frac{RT}{4F} \ln\left( \frac{p_{O_2,b}}{p_{O_2,TPB}} \right) , \qquad (3)$$

Where the subscripts b, TPB represent the bulk and three-phase interface, respectively, and $p_i$ is the partial pressure of the component $i$ at the corresponding position, which can be taken as ($H_2$, b), ($H_2O$, TPB), ($H_2$, TPB), ($H_2O$, b) positions, respectively.

The exchange current density $i_0$ can be calculated for[23] according to

$$i_{0,a}^{H_2} = \gamma_{a,H_2} \left(\frac{p_{H_2}}{p_{ref}}\right)^m \exp\left(-\frac{E_{act,a}}{RT}\right) , \qquad (4)$$

$$i_{0,c}^{O_2} = \gamma_{a,O_2} \left(\frac{p_{O_2}}{p_{ref}}\right)^m \exp\left(-\frac{E_{act,c}}{RT}\right) , \qquad (5)$$

Where $m$, $k$ are reaction orders respectively, $\gamma_{a,H_2}$ and $\gamma_{c,O_2}$ are adjustable parameters, $p_{ref}$ is the reference pressure, which is usually taken as 1 atm (1 atm = 1.01 × 10$^5$ Pa), $E_{act,a}$ is the activation energy of hydrogen reaction at the anode, and $E_{act,c}$ is the activation energy of oxygen reaction at the cathode.

2.2 Gas flow model

During the operation of SOFC, the fuel gas and oxidizing gas can diffuse and flow freely between the electrode and the channel. In the numerical simulation, the flow of fluid in the porous electrode and the gas channel needs to be simulated, but the governing equations are different for the porous electrode and the gas channel.

For the momentum governing equations in the anode channel and cathode channel, the Navier-Stokes equation[24] is used to describe the equation of motion for momentum conservation of viscous incompressible fluid as follows:

$$\nabla(\rho \cdot \vec{u}) = 0 , \qquad (6)$$

$$\rho(\vec{u} \cdot \nabla) = \nabla \cdot \left[-p \cdot \mathbf{I} + \mathbf{\iota}\right] + \vec{Q} , \qquad (7)$$

Where $\nabla$ is the Laplace operator, and $\rho$ is the density of the mixed gas, which is provided by the material diffusion equation; The $\mathbf{u}$ is the velocity vector; The $p$ is the gas pressure; $\mathbf{I}$ is the unit tensor, and $\mathbf{\iota}$ is the viscous stress tensor, which is related to the gas velocity, dynamic viscosity, and temperature; $\mathbf{Q}$ is the volume force vector.

The gas flow in the porous electrode is described by the Brinkman[25] equation, which can be used to solve the fluid velocity of the porous medium in laminar flow:

$$\frac{\rho}{\varepsilon}(\vec{u} \cdot \nabla)\vec{u} = -\varepsilon \nabla \left[\mu\left(\boldsymbol{\delta} - \frac{2}{3}\nabla \cdot \vec{u}\right)\right] + \varepsilon \Delta P , \qquad (8)$$

Where $\varepsilon$ is an auxiliary parameter, $\mathbf{u}$ is the velocity vector, $\boldsymbol{\delta}$ is the viscosity tensor, $\Delta P$ is the momentum loss of the fluid after passing through the porous medium, and $\mu$ is the viscosity coefficient of the fluid.

2.3 Material diffusion model

The mass transfer in the electrode mainly occurs in the gas phase, and the mass transfer in the chemical reaction is dominated by diffusion and convection. Convection mainly occurs in the gas channel, and diffusion mainly occurs inside the porous electrodes of the cathode and anode. The mass conservation equation of each gas substance can be expressed as

$$\rho(u \cdot \nabla)w_i + \nabla \vec{J}_i = R_i \quad , \tag{9}$$

Where $w_i$ is the mass fraction of each gas, $J_i$ is the flux vector associated with $w_i$, and $R_i$ is an additional source term for material production and consumption $R_i = 0$.

In porous electrodes of cathode and anode, the extra source term $R_i$ for matter production and consumption is:

$$R_i = \frac{c_i i_v}{nF} \quad , \tag{10}$$

Where $c_i$ is the stoichiometric coefficient of each substance, $i_v$ is the volume current density of the electrochemical reaction, and $n$ is the number of participating electrons in the electrochemical reaction.

The mixed gas density $\rho$ is

$$\rho = \frac{p \sum x_i M_i}{RT} \quad , \tag{11}$$

Where $x_i$ is the mole fraction of the substance and $M_i$ is the molar mass of the substance.

Maxwell-Stefan is a theoretical model describing the diffusion process of multi-component system in thermodynamics. Based on this model[26], the mass diffusion flux $J_i$ can be expressed as

$$J_i = -\left( \rho w_i \sum_k D_{ik} d_k + D_i^T \frac{\nabla T}{T} \right) \quad , \tag{12}$$

Where $D_i^T$ is the thermal diffusivity, $D_{ik}$ is the Maxwell-Stefan diffusivity, and $\nabla T$ is the temperature gradient.

## 3. SOFC finite element model

COMSOL Multiphysics software was used to study the electrochemical reaction of SOFC model under the coupling of electrochemical model, gas flow model and species diffusion model. The numerical simulation of SOFC is a process of solving complex multi-physical field coupling equations. In order to save the calculation time while ensuring the calculation accuracy, it is necessary to make corresponding assumptions and simplifications for the complex system based on the actual operation of SOFC. Firstly, the whole reaction process is assumed to be steady state, and the start-up and shutdown instants of SOFC are not considered; Secondly, the mixed gas is composed of ideal gas, and the reaction gas is considered to flow in a smooth flow channel with the same cross section; Third, because the chemical reaction layers of the anode and cathode are very thin, it is assumed that the redox reaction occurs at the interface between the electrode and the electrolyte; Fourth, the porous electrode material is uniform and isotropic.

Anode-supported planar SOFC is mainly composed of anode connector, anode channel, anode diffusion layer, anode, electrolyte, cathode, cathode diffusion layer, cathode channel and cathode connector. The channels are distributed in the form of forward/reverse flow or cross flow respectively, and their geometric structure is shown in Fig. 1(a),(b). The cathode channel and anode channel of SOFC are composed of 8 channels, and the geometric parameters of the model are shown in the Tab. 1. The simulation in this paper focuses on the electrochemical performance of the battery under the coupling of flow pattern and flow rate. The smaller size of the battery model is helpful to observe and analyze more clearly, and the detailed results can be obtained more quickly in the simulation process. Based on the physical model used in reference [27], a multi-physical field coupling model of anode-supported planar SOFC with a reaction area of 20 mm × 20 mm was established by COMSOL Multiphysics software. The finite element models of forward/reverse flow and cross flow are shown in Fig. 1(c),(d).

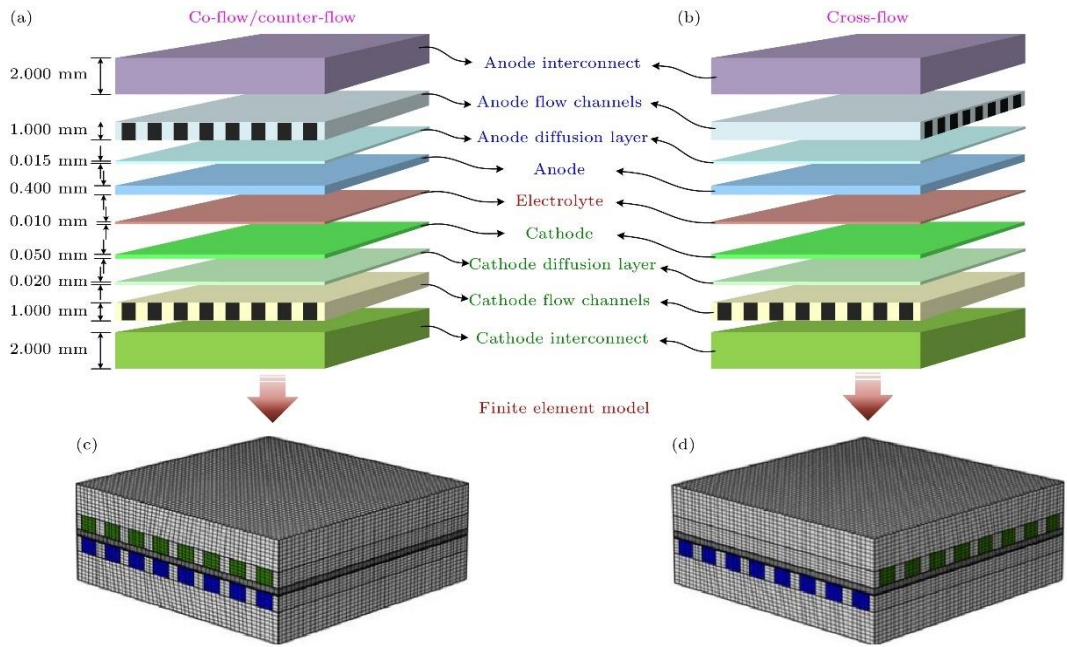

**Figure 1.** Geometry structure and finite element model of SOFC: (a) Geometry structure of co-flow/counter-flow patterns; (b) geometry structure of cross-flow pattern; (c) finite element model of co-flow/counter-flow patterns; (d) finite element model of cross-flow pattern.

**Table 1.** Geometric parameters of SOFC single-cell model.

| Geometric parameter | Numerical value |
|---|---|
| Battery length/mm | 20.000 |
| Cell width/mm | 20.000 |
| Flow channel height/mm | 1.000 |
| Flow channel width/mm | 1.500 |
| Rid width/mm | 1.000 |
| Anode and Cathode Connector Tickness/mm | 2.000 |
| Thickness of anode diffusion layer/mm | 0.015 |
| Anode thickness/mm | 0.400 |
| Cathode diffusion layer thickness/mm | 0.020 |
| Cathode thickness/mm | 0.050 |
| Electrolyte thickness/mm | 0.010 |

In this study, the top surface of the metal interconnect on the anode side is defined as ground, and the anode equilibrium potential is 0 V, while the bottom surface of the metal interconnect on the cathode side is defined as the cell voltage, and the potential is set to $V_{cell}$ = 0.5 V under actual working conditions. The gas inlet is assumed to be laminar and the mean velocity is specified as the boundary condition, the anode inlet velocity is set in the range of 0.5 — 2.5 m/s, the cathode inlet velocity is set in the range of 0.5 — 1.0 m/s, the gas outlet is the pressure boundary, the atmospheric pressure is set at 1 atm (1 atm = 1.013 × 10$^5$ Pa), and the backflow is suppressed. In addition, the inlet gas of the anode channel is defined as hydrogen with a mole fraction of 0.95 and water vapor with a mole fraction of 0.05, while the inlet gas of the cathode channel is defined as oxygen with a molar fraction of 0.21 and nitrogen with a molar fraction of 0.79. The initial temperature, boundary temperature and gas inlet temperature are all set to 1073 K. The boundary condition is set as free boundary, assuming that the battery is not loaded by external force, and the components of the battery can deform freely.

Yttria stabilized zirconia (YSZ)[28] was used as the electrolyte material, nickel-Yttria stabilized zirconia (Ni-YSZ) was used as the anode material and anode support material, and strontium-doped lanthanum manganate (LSM) was used for the cathode material. The electrochemical model parameters, gas flow model parameters and species diffusion model parameters of SOFC are shown in Tab. 2 —Tab. 4, respectively[29–31].

Table 2. Electrochemical model parameters[28,29].

| Parameter | Numerical value |
|---|---|
| Anode equilibrium potential/V | 0 |
| Battery operating voltage/V | 0.5 |
| Anode exchange current density/(A/m$^2$) | 5 |
| Cathode exchange current density/(A/m$^2$) | 2 |

| Parameter | Numerical value |
| --- | --- |
| Anode active specific surface area/m$^{-1}$ | $1\times 10^5$ |
| Cathode active specific surface area/m$^{-1}$ | $1\times 10^5$ |
| Electrolyte conductivity/(S/m) | 5 |
| Anode conductivity/ (S/m) | 1000 |
| Cathode conductivity/ (S/m) | 1000 |
| Conductivity of anode diffusion layer/(S/m) | $8.5\times 10^5$ |
| Conductivity of cathode diffusion layer/(S/m) | 7700 |
| Conductivity of current collector/(S/m) | $1.4\times 10^6$ |

Table 3. Gas flow model parameters[23,30].

| Parameter | Numerical value |
| --- | --- |
| Atmospheric pressure/ATM | 1 |
| Anode permeability/m$^2$ | $1\times 10^{-12}$ |
| Cathode permeability/m$^2$ | $1\times 10^{-12}$ |
| Anode porosity | 0.3 |
| Cathode porosity | 0.3 |
| Anode channel gas inlet velocity/ (m·s$^{-1}$) | 0.5 ~2.5 |
| Gas inlet velocity of cathode channel/ (m·s$^{-1}$) | 1.0 ~ 0.5 |

Table 4. Material diffusion model parameters[28,29].

| Parameter | Numerical value |
| --- | --- |
| Reference diffusivity/ (m$^2$· s$^{-1}$) | $3.16\times 10^{-8}$ |
| Fuel gas pore volume fraction/% | 40% |
| Oxidized gas pore volume fraction/% | 40% |
| Molar mass of hydrogen/ (g·mol$^{-1}$) | 2 |
| Molar mass of oxygen/ (g·mol$^{-1}$) | 32 |
| Molar mass of water vapor/(g·mol$^{-1}$) | 18 |
| Molar mass of nitrogen/ (g·mol$^{-1}$) | 28 |

## 4. Calculation results and analysis

4.1 Finite element model verification

The finite element model for the study of the electrochemical performance of SOFC under the influence of gas flow and voltage is based on certain simplifications and assumptions, so it is necessary to verify the accuracy and precision of the finite element model. In this study, the accuracy and precision of the established SOFC finite element model were verified by establishing a finite element model with the same cell size and relevant physical parameters as those studied in reference [27], setting the fuel gas composition as 58.2% $H_2$, 3% $H_2O$, 19.4% CO, 19.4% $CO_2$ according to reference [27], and comparing the polarization curve and the distribution of gas composition under the same voltage. The polarization contrast curves of the cell are shown in the Fig. 2, and the results show that the two polarization curves are in good agreement, especially in the working voltage range of 0.9 — 0.4 V. The component distribution of hydrogen, oxygen and water vapor in the cell at the same voltage is shown in Fig. 3. It can be seen from Fig. 3 that the concentration of each component is consistent with the results obtained in reference [27].

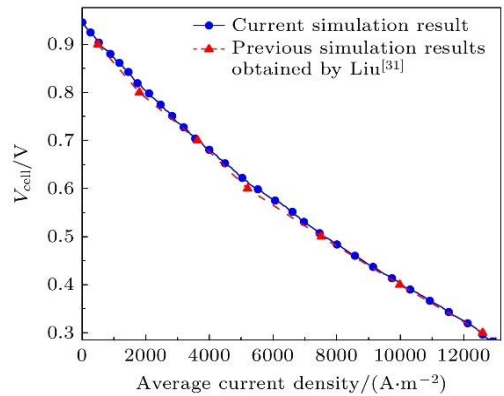

**Figure 2.** Comparison of the polarization curves between the results of current finite element model and the results obtained in Ref. [27].

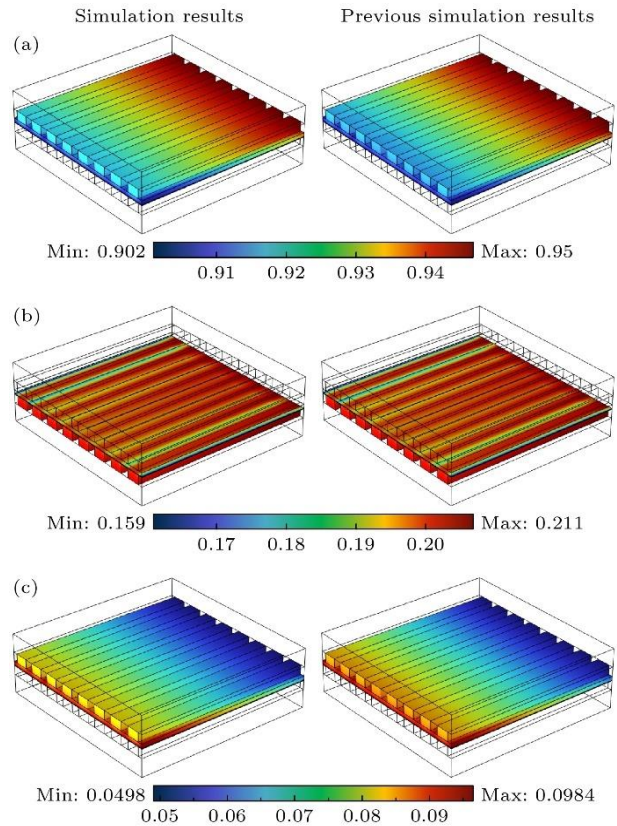

**Figure 3.** Comparison of the gas components distribution between the results of current finite element model and the results obtained by Ref. [27]: (a) Hydrogen mole fraction; (b) oxygen mole fraction; (c) water vapor mole fraction.

4.2 Effect of Operating Voltage on Electrochemical Performance of SOFC

Voltage, as one of the key parameters of SOFC, directly affects the driving force of electrode reaction and the output performance of the cell. By exploring the variation of SOFC composition distribution and electrolyte current density under different voltages, the performance boundary of the cell can be determined, and the optimal operating point can be found to achieve the highest energy conversion efficiency. This also helps to reveal the reaction kinetics mechanism and mass transfer process limiting factors in the battery, and provides a strong basis for the structural optimization and material improvement of the battery.

The current density in the electrolyte depends on the electrochemical reaction rate, which is affected by the voltage, the length of the three-phase boundaries (TPB), and the concentration of the reactant gas. The current density distribution of SOFC electrolyte at different operating voltages is given by Fig. 4. When the gas flow pattern is co-current, the Fig. 4(a) is the graph of the maximum and minimum values of electrolyte current density at different voltages. The Fig. 4 show that the electrolyte current density decreases with the increase of voltage, and the difference between the maximum and minimum values also decreases gradually. When the cell voltage is 0. 9 V, the current density is at a relatively low level. From the Fig. 4(b), it can be seen that the color change is relatively gentle, which indicates that the current density changes slightly inside the cell. At this time, the maximum current density is about 209 A/m$^2$, and the minimum is about 207 A/m$^2$. When the cell voltage is reduced to 0. 6 V, the current density is higher than that at 0. 9 V, and the color change in the Fig. 4(c) is more significant than that in the Fig. 4(b). At this voltage, the maximum current density is about 1290 A/m$^2$, and the minimum is about 1230 A/m$^2$. When the cell voltage is further reduced to 0. 3 V, the current density continues to increase, and the color change in Fig. 4(d) becomes more intense. At this time, the maximum current density is about 6450 A/m$^2$, and the minimum current density is about 4750 A/m$^2$. With the cell voltage gradually decreasing from 0. 9 V to 0. 3 V, the electrolyte current density increases in the SOFC downstream case. At the same time, the inhomogeneity of current density distribution in the cell increases with the decrease of voltage, that is, the variation of current density increases. This phenomenon indicates that the electrochemical reaction inside the cell is more intense at lower voltage, resulting in a more significant change in current density.

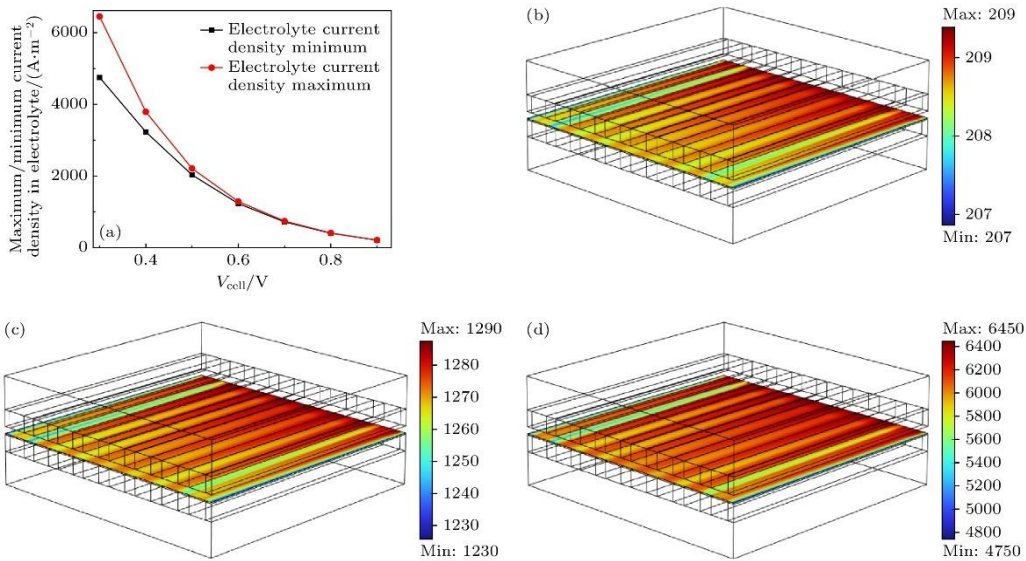

**Figure 4.** SOFC electrolyte current density distribution in the case of downstream: (a) Plot of maximum and minimum electrolyte current density at different voltages; (b) electrolyte current density cloud at voltage of 0.9 V; (c) electrolyte current density cloud at voltage of 0.6 V; (d) electrolyte current density cloud at voltage of 0.3 V.

When the gas flow pattern is downstream, considering the chemical reaction kinetics, gas distribution uniformity and other factors, the gas flow rate of the cathode channel is set to 1 m/s, and the gas flow rate of the anode channel is set to 0.4 m/s. The range of hydrogen mole fraction of the contact surface between the anode electrode and the channel and oxygen mole fraction of the contact surface between the cathode electrode and the channel under different working voltages is shown in Fig. 5. It can be seen from the figure that the range of the mole fraction of each component increases gradually with the decrease of the cell voltage. When the cell voltage is 0. 9 V, the mole fraction ranges of hydrogen and oxygen are 0. 004 and 0. 005, respectively. When the cell voltage is 0. 6 V, the mole fraction ranges of hydrogen and oxygen are 0. 024 and 0. 03, respectively. When the cell voltages is 0.

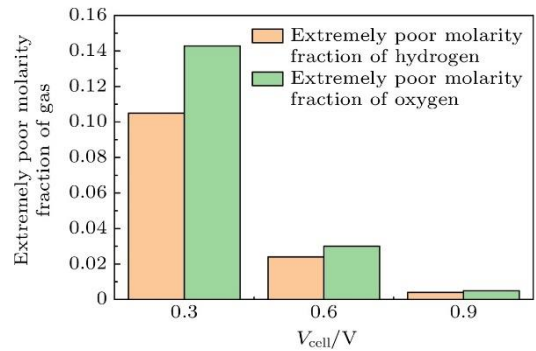

**Figure 5.** The extremes of the molar fraction of hydrogen, oxygen at different voltages.

4.3 Effect of gas flow pattern on electrochemical performance of SOFC

Fig. 6 is the polarization curve and power density curve of SOFC under different flow patterns. It can be seen from the Fig. 6(a) that the average current density increases with the decrease of voltage. Under the condition of high voltage, different flow patterns have little effect on the electrochemical performance of the fuel cell; Under the condition of low voltage, the electrochemical performance of the cross flow pattern is obviously better than that of the other two flow patterns. It can be seen from the Fig. 6(b) that the average cell power density increases with the increase of the average current density, and the average cell power density of the cross flow is the highest in the whole current density range, followed by the counter flow, and the forward flow is the lowest. Compared with the forward flow and the reverse flow, the maximum output power of the cross flow is increased by 22.4% and 4.9%, respectively.

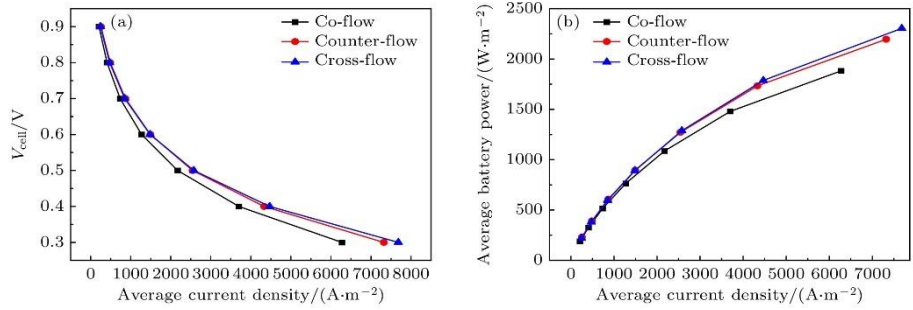

**Figure 6.** Polarization curves and power density curves of SOFC under different flow modes: (a) Polarization curves; (b) power density curves.

The Fig. 7 and Fig. 8 give the component distribution of hydrogen and oxygen under different flow patterns at different voltages. The analysis shows that the decrease of cell voltage will increase the mole fraction gradient of gas components in different flow patterns, which means that the electrochemical reaction rate is faster. By comparing the gas component distribution in different channels under different voltages, it is found that the change of gas mole fraction is the most uniform under the co-current flow pattern. Under the condition of low voltage, the gas consumption of countercurrent flow pattern is more uneven. The cross-flow flow pattern has some advantages in oxygen supply under the ribs, although the electrochemical reaction is more uneven at low voltage.

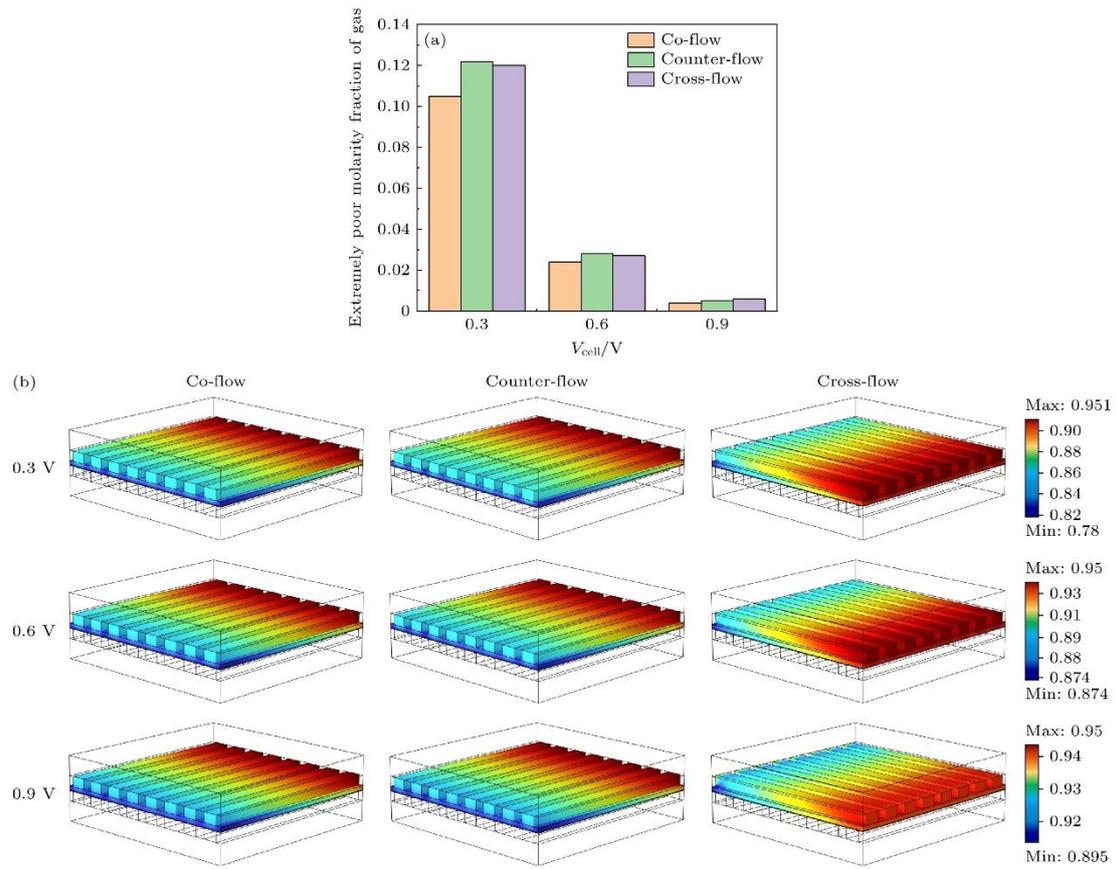

**Figure 7.** Distribution of hydrogen in different flow channels under different voltages: (a) Histogram of the range of hydrogen mole fraction; (b) hydrogen molarity fraction distribution.

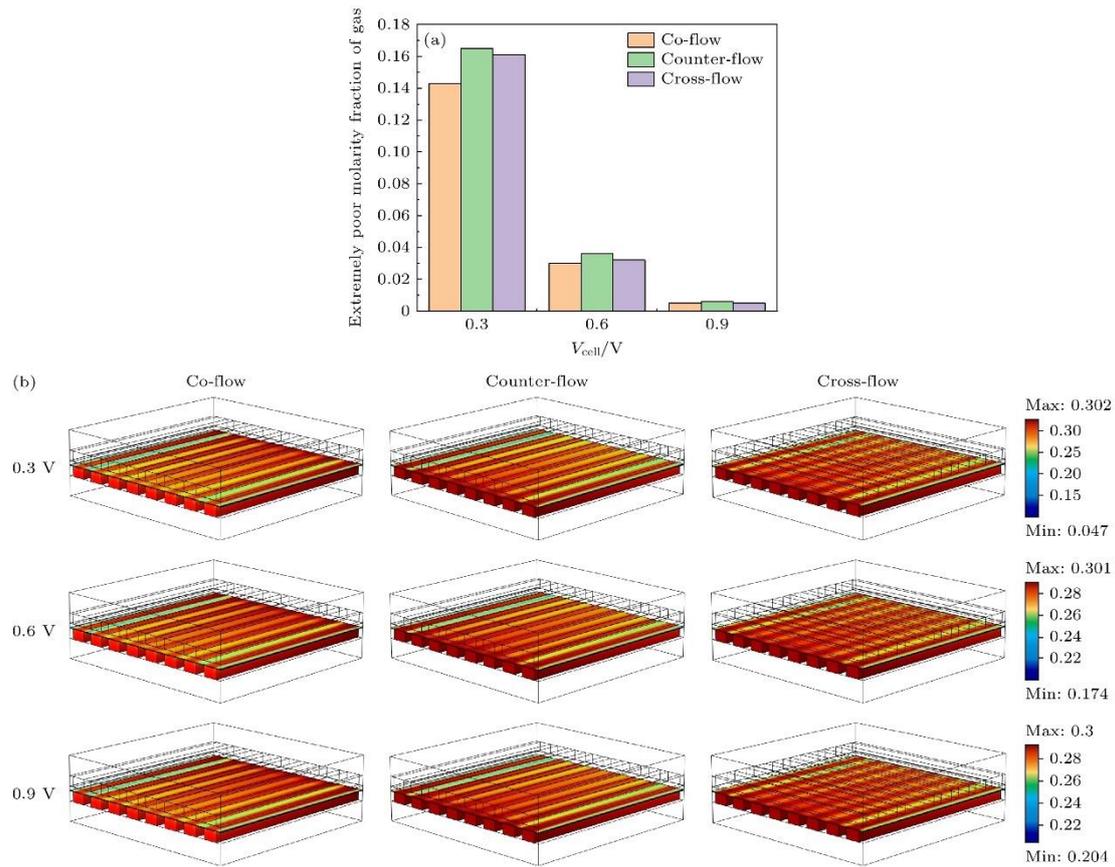

**Figure 8.** Distribution of oxygen in different flow channels under different voltages: (a) Histogram of the variation range of oxygen mole fraction; (b) contour map of oxygen mole fraction distribution.

The Fig. 7(a) is the hydrogen mole fraction range histogram of different flow patterns at different voltages, and the Fig. 7(b) is the hydrogen mole fraction distribution nephogram at the anode side of different flow patterns at different voltages. It can be seen from the Fig. 7(a) that when the cell voltage is 0. 3 V, the range difference is 0. 105 in the forward flow mode, 0.122 in the reverse flow mode, and 0. 12 in the cross flow mode. This indicates that the minimum hydrogen mole fraction in the reverse flow and cross flow mode is higher than that in the forward flow mode at a lower voltage, and the hydrogen concentration distribution is more likely to have a very low value than that in the parallel flow mode. When the cell voltage is 0. 6 V, the difference between the extreme values of the three flow patterns narrows, indicating that the influence of different flow patterns on the minimum value of hydrogen mole fraction decreases with the increase of voltage. When the cell voltage is 0. 9 V, the range values of the three flow patterns are at a very low level, indicating that different flow patterns have little effect on the minimum value of hydrogen mole fraction at high voltage.

From the Fig. 7(b), it can be seen that the hydrogen mole fraction decreases gradually along the flow direction, and the distribution of the mole fraction of each component on the anode side is similar under the three flow patterns because the flow direction on the anode side is unchanged. When the voltage is 0. 6 V, the range of hydrogen mole fraction is 0. 926 — 0.95 in the co-current flow pattern, 0. 922 — 0.95 in the counter-current flow pattern, and 0. 874 — 0.901 in the cross-current flow pattern. It can be seen that the range of hydrogen mole fraction on the anode side is the largest in the cross-current flow pattern and the smallest in the co-current flow pattern. It indicates that compared with the other two flow patterns, the electrochemical properties of the cross flow pattern.

The Fig. 8(a) is the oxygen mole fraction range histogram of different flow patterns at different voltages, and the Fig. 8(b) is the oxygen mole fraction distribution nephogram at the anode side of different flow patterns at different voltages. It can be seen from the Fig. 8(a) that when the cell voltage increases from 0. 3 V to 0. 6 V, the oxygen range value decreases from 0. 1428 to 0. 03 when the SOFC is in the forward flow condition, and when the cell voltage is 0. 9 V, the range value is almost 0, which indicates that with the increase of the voltage, the oxygen concentration distribution under the forward flow condition is more uniform, and the occurrence of very low oxygen mole fraction decreases. When the SOFC is in counterflow and crossflow conditions, the range of oxygen mole fraction decreases and the oxygen concentration distribution tends to be stable with the increase of cell voltage.

From the Fig. 8(b), it can be seen that when the voltage is 0. 6 V, the oxygen mole fraction of the co-current flow pattern varies from 0. 18 to 0. 21, the oxygen mole fraction of the counter-current flow pattern varies from 0. 174 to 0. 21, and the oxygen mole concentration of the cross-current flow pattern ranges from 0. 268 to 0.

The gas velocity profiles in the anode channel and cathode channel of SOFC at cell voltage of 0.9 V and 0.5 V are given by Fig. 9 and Fig. 10, respectively. The analysis shows that the decrease of cell voltage will make the gas velocity distribution in the flow channel more uneven, which means that the electrochemical reaction rate increases. At low voltage, the gas velocity distribution of counterflow and crossflow changes more significantly, which has a greater impact on the electrochemical performance, while the counterflow is relatively uniform.

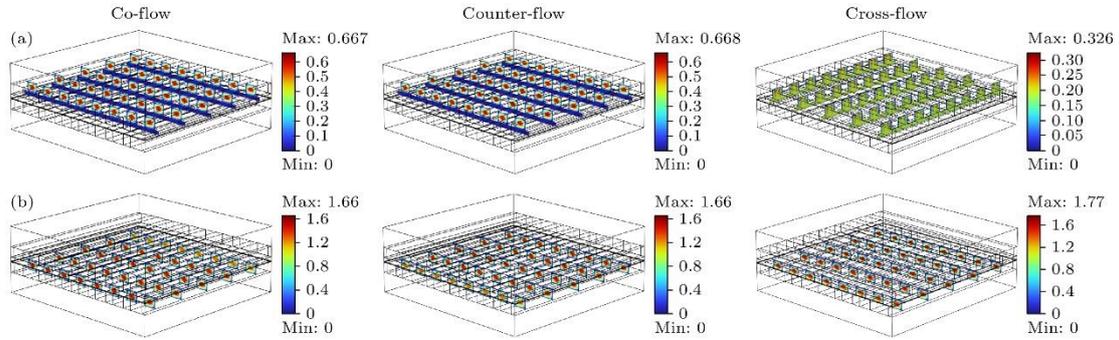

**Figure 9.** Velocity distribution in the flow channel when the cell voltage is 0.9 V: (a) Velocity distribution in the anode flow channel; (b) velocity distribution in the cathode flow channel.

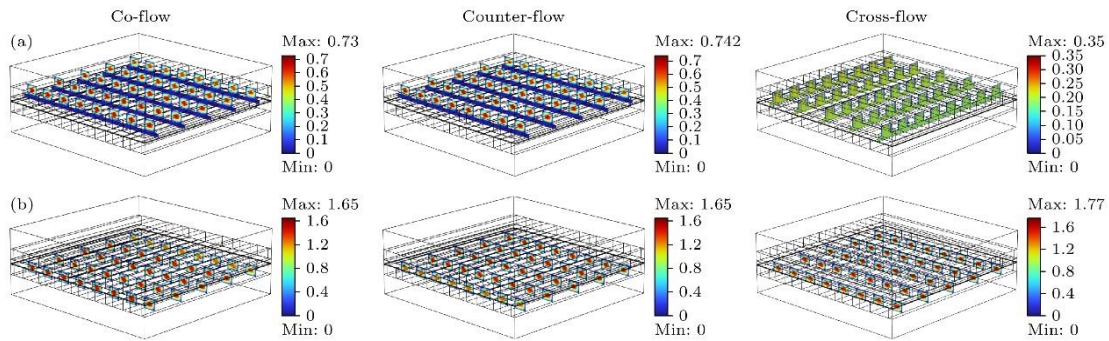

**Figure 10.** Velocity distribution in the flow channel when the cell voltage is 0.5 V: (a) Velocity distribution in the anode flow channel; (b) velocity distribution in the cathode flow channel.

4.4 Ffect of gas velocity on electrochemical performance of SOFC with different flow regime

Considering the sufficient degree of chemical reaction and other factors, the gas inlet rate of the cathode channel was set at 2 m/s, and the gas flow rates of the anode channel were set at 0. 1,0.2,0.3,0.4 and 0. 5 m/s, respectively. The polarization curves of different flow patterns SOFC anode channel gas at different flow rates, forward flow, reverse flow and cross flow are shown in Fig. 11. The results show that under the same voltage, the average current density of the cross-flow type is the largest, followed by the counter-flow type, and the average current density of the forward flow type is the smallest. The average current density of the three flow patterns increases with the increase of the gas flow rate in the anode channel. It can be clearly seen that the average current density of the cross-flow pattern changes in the largest range with the increase of the gas flow rate in the anode channel, from 7158 A/m² to 7783 A/m², an increase of 8.7%.

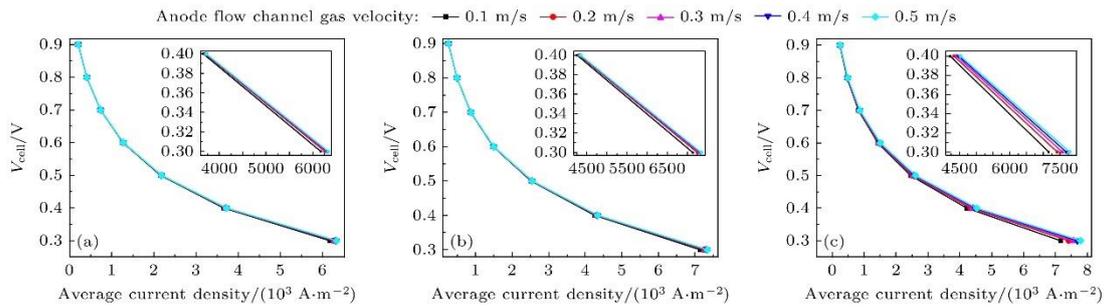

**Figure 11.** SOFC polarization curves under different airflow velocities of anode channel gas: (a) Co-flow; (b) counter-flow; (c) cross-flow.

The power density curves of different flow patterns SOFC anode channel gas at different flow rates, forward flow, reverse flow and cross flow are shown in Fig. 12. The results show that the output power density of the cell increases with the increase of the gas flow rate in the anode channel. The reason for this phenomenon is that with the increase of the gas flow rate in the anode channel, the amount of hydrogen gas per unit time increases, which increases the hydrogen concentration in the anode channel, accelerates the electrochemical reaction rate, and reduces the loss caused by concentration polarization. The electrochemical performance of the cell can be improved by increasing the gas flow rate in the anode channel, and the maximum output power density increases from 1850 W/m² to 1898 W/m² under the condition of forward flow, which is improved by 2.5%. The maximum output power density in countercurrent flow increases from 2140 W/m² to 2201 W/m², an increase of 2.8%. In the case of cross flow, the maximum output power density increases from 2147 W/m² to 2335 W/m², an increase of 8.7%. Through the study of the relationship between the gas flow rate in the anode channel and the cell polarization curve and power density curve under different flow patterns, it can be clearly seen that the gas flow rate in the anode channel has a vital impact on the cell performance. The increase of flow rate can effectively improve the maximum output power density of the cell. At the same time, compared with the other two flow patterns, the maximum output power density of the cross-flow pattern is increased more significantly under the same change of gas flow rate in the anode channel, which indicates that the cross-flow pattern has unique advantages in gas mass transfer and electrochemical reaction synergy, and has more potential in optimizing the performance of the cell.

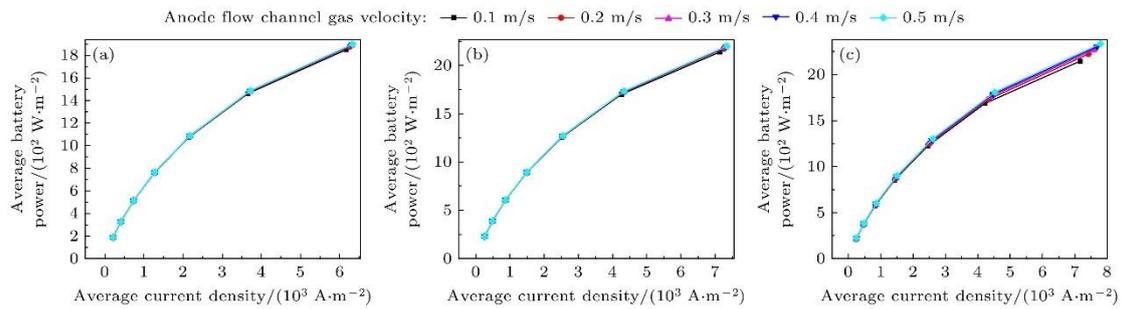

**Figure 12.** SOFC power density curves at different airflow velocities of anode channel gases: (a) Co-flow; (b) counter-flow; (c) cross-flow.

When the gas flow rate in the anode channel is set at 0. 4 m/s and the gas flow rate in the cathode channel is 0. 5, 1, 1. 5, 2 and 2. 5 m/s, the polarization curves of the SOFC cells with different flow patterns, such as forward flow, reverse flow and cross flow, are shown in Fig. 13. The results show that under the same voltage condition, the average current density of the cross-flow type is in the leading position, the average current density of the counter-flow type is in the second place, and the average current den- sity of the co-flow type is the smallest. With the increase of gas flow rate in the cathode channel, the average current density of the three flow patterns also increases gradually, but compared with the change of gas flow rate in the anode channel, the effect of changing the gas velocity in the cathode channel on the average current density of the cross-flow pattern is significantly reduced, and the maximum value is increased from 7620 A/m² to 7725 A/m², which is increased by 1. 3%. However, it can be seen from the numerical values that when the gas flow rate in the cathode channel is changed, the average current density of different flow patterns under the same voltage is higher than that when the gas flow rate in the anode channel is changed.

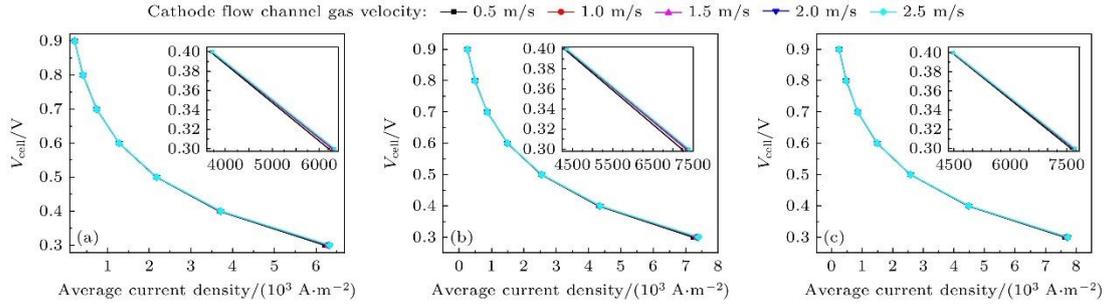

**Figure 13.** SOFC polarization curves under different airflow velocities of cathode flow channel gases: (a) Co-flow; (b) counter-flow; (c) cross-flow.

The power density curves of SOFC with different flow patterns, such as forward flow, reverse flow and cross flow, are shown in Fig. 14. The results show that with the increase of the gas flow rate in the cathode channel, the output power density of the cell shows a significant increase. The reason for this phenomenon is that with the increase of the gas flow rate in the cathode channel, the amount of oxygen gas per unit time increases, which leads to the increase of oxygen concentration in the anode channel and accelerates the electrochemical reaction rate, thus improving the electrochemical performance. However, as the gas flow rate in the cathode channel continues to increase, the increase in the output power density decreases, because the electrochemical reaction of the cell gradually reaches saturation, and the effect of the increase in oxygen concentration on the increase in the electrochemical reaction rate weakens, so the output power density decreases. The maximum output power density in forward flow increases from 1860 W/m$^2$ to 1898 W/m$^2$, an increase of 2%. In the case of countercurrent, the maximum output power density increases from 2165 W/m$^2$ to 2217 W/m$^2$, an increase of 2. 4%. The peak output power density in the cross-flow case increases from 2286 W/m$^2$ to 2317 W/m$^2$, an increase of 1. 3%. In the study of the factors affecting the cell performance, it was found that for the cross-flow type, the growth rate of the maximum output power density was slower when the gas flow rate in the cathode channel was changed than when the gas flow rate in the anode channel was changed. It is worth noting that increasing the gas flow rate in the cathode channel can still effectively improve the maximum output power density of the cell. This phenomenon indicates that the gas flow rate in the cathode channel is not as effective as that in the anode channel in improving the cell performance, but it is still an important controllable parameter in optimizing the cell performance.

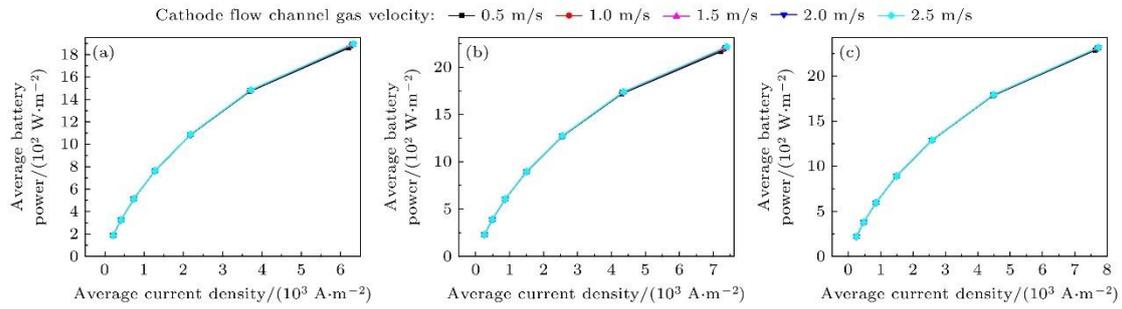

**Figure 14.** SOFC power density curves at different airflow velocities of cathode channel gases: (a) Co-flow; (b) counter-flow; (c) cross-flow.

## 5. Conclusion

In this paper, COMSOL Multiphysics finite element simulation software was used to construct a three-dimensional multi-field coupling model of SOFC, and the electrochemical performance of SOFC under three flow patterns of co-current, counter-current and cross-current was studied. In this paper, the effects of different working conditions on the electrochemical performance of SOFC are analyzed in depth through the model, and the main conclusions are as follows.

1) Under the gas co-current arrangement, when the gas flow rate in the anode channel is 1 m/s and the gas flow rate in the cathode channel is 0. 4 m/s, the electrochemical reaction is accelerated and the mole fraction gradient of anode hydrogen and cathode oxygen increases when the voltage is reduced from 0. 9 V to 0. 3 V. At low voltage, the uneven distribution of electrolyte current density in SOFC is aggravated, and the electrochemical reaction is intense.

2) According to the polarization curve and power density curve, it can be seen that the high voltage has little effect on the SOFC of different flow patterns, and the cross flow performance is excellent at low voltage. Compared with forward flow and reverse flow, the maximum output power of cross flow is increased by 22.4% and 4.9%, respectively. The change of gas mole fraction is the most uniform in forward flow, and the gas consumption is more uneven in reverse flow at low voltage. Although the electrochemical reaction is more uneven in cross flow at low voltage, it has an advantage in oxygen supply under the rib.

3) The increase of gas flow rate in the anode channel can effectively improve the maximum output power density of the cell, and the increase is more significant in the cross-flow mode under the same flow rate change. The increase of gas flow rate in the cathode channel can also increase the output power density of the cell, but with the increase of gas flow rate, the increase of output power density decreases due to the saturation of electrochemical reaction. Under different flow patterns, concurrent flow, countercurrent flow and cross flow all show the same variation trend of power density. At the same time, the effects of changing the gas flow rate in the anode and cathode channels on the maximum output power density of the cross-flow fuel cell were compared, and it was found that the growth rate slowed down when changing the gas flow rate in the cathode channel, but it was still an important adjustable parameter to optimize the performance of the fuel cell.

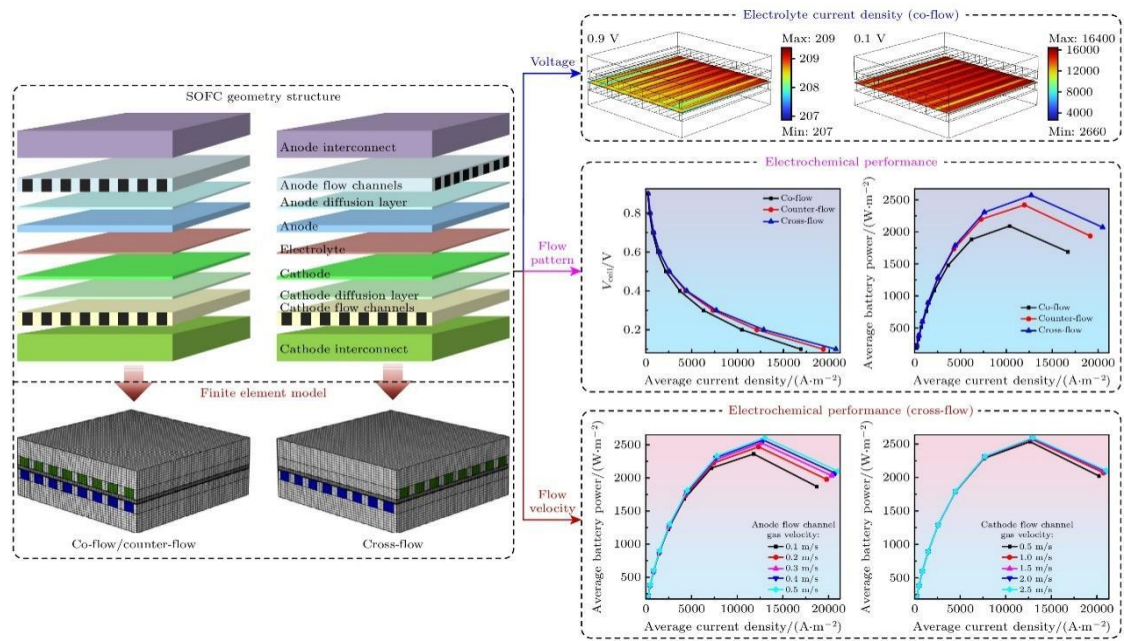